\documentclass{amsart}
\usepackage{verbatim} 
\usepackage{amsfonts,amssymb,amsmath}

\newtheorem{thm}{Theorem}
\newtheorem{note}{Note}
\newtheorem{Def-Not}{Definition--Notation}
\newtheorem{prop}{Proposition}

\begin{document}

\newtheorem{definition}{Definition}[section]
\newtheorem{proposition}{Proposition}[section]
\newtheorem{remark}{Remark}[section]
\newtheorem{Restriction}{Restriction}
\newtheorem{Remark}{Remark}[section]
\newtheorem{corollary}{Corollary}[section]
\newtheorem{Theorem}{Theorem}[section]
\newtheorem{theorem}{Theorem}[section]
\newcommand{\const}{\mbox{const.}}
\newtheorem{thrm}{Theorem}[section]
 \newtheorem{lem}{Lemma}[section]
\newtheorem{dfntn}{Definition}[section]
\newtheorem{sttmnt}{Proposition}[section]
\newtheorem{rmrk}{Remark}[section]
\renewcommand{\thepage}{ \arabic{page}}
\newtheorem{Lemma}{Lemma}[section]

\def\CC{\mathbb{C}}
\def\NN{\mathbb{N}}
\def\CP1{\mathbb{C}{\mathbf P}^1}
\def\ZZ{\mathbb{Z}}
\def\RR{\mathbb{R}}
\def\OOO{\mathbf{O}}
\def\cL{{\mathcal  L}}
\def\gl1{\operatorname{gl}(1, \CC)}
\def\GL1{\operatorname{GL}(1, \CC)}
\def\SL2C{\operatorname{SL}(2,\mathbb{C})}
\def\sl2C{\operatorname{sl}(2,\mathbb{C})}
\def\glmC{\operatorname{gl}(m,\mathbb{C})}
\def\GLmC{\operatorname{GL}(m,\mathbb{C})}

\def\glm{\operatorname{gl}(m)}
\def\GLm{\operatorname{GL}(m)}
\def\SLmC{\operatorname{SL}(m,\mathbb{C})}
\def\slmC{\operatorname{sl}(m,\mathbb{C})}

\def\Res{\operatorname{Res}}
\def\Ad{\operatorname{ad}}
\def\Tr{\operatorname{tr }}
\def\sn{\operatorname{sn }}
\def\cn{\operatorname{cn }}
\def\dn{\operatorname{dn }}
\def\const{ \mathfrak{const}}

\title{Rational symplectic coordinates \\ on the space of Fuchs equations\\
$m\times m$-case}%
\address{PDMI, Fontanka 27, St.Petersburg, 191023, Russia}%
\email{mbabich@pdmi.ras.ru}%
\author{Mikhail V. Babich}
\begin{abstract}
A method of  constructing of Darboux
 coordinates on a space that is a symplectic
   reduction with respect to a diagonal action of $\operatorname{GL}(m,\mathbb{C})$
   on a Cartesian product of $N$ orbits of coadjoint representation
   of  $\operatorname{GL}(m,\mathbb{C})$ is presented.
   The method gives
   an explicit symplectic birational isomorphism between the reduced space on the one hand
   and a Cartesian product of $N-3$ coadjoint orbits of dimension $m(m-1)$ on an orbit of
    dimension $(m-1)(m-2)$ on the other hand.
   In a generic case of the
   diagonalizable matrices it gives just the isomorphism
   that is birational and symplectic between some open,
    in a Zariski topology, domain of the reduced space and the
 Cartesian product of the orbits in question.

   The method is based on a Gauss decomposition
    of a matrix on a product of upper-triangular, lower-triangular and diagonal matrices.
   It works uniformly for the orbits formed by diagonalizable or not-diagonalizable
   matrices.   It is elaborated for the
   orbits of maximal dimension that is $m(m-1)$.

   \end{abstract}
\maketitle

\setcounter{section}{-1}
\section{Introduction}
A set of Fuchs equations\footnote{Ground field is $\CC$,
$\operatorname{GL}(m)$ means $\operatorname{GL}(m,\mathbb{C})$ etc.}
\begin{equation*}
   \frac{d}{d\lambda}\Psi
=\sum_{n=1}^{N}\frac{A^{(n)}}{\lambda-\lambda_n}\Psi, \ \
\Psi\in \GLm, \ \ A^{(n)}\in \glm
\end{equation*}
may be considered as a submanifold $\sum_{n=1}^NA^{(n)}=0$ of the space
$\glm \times \dots \times \glm$ that has  a natural Poisson structure  coming from $\glm$.

Some important problems like a problem of isomonodromic deformations
may be restricted on a ``symplectic leaf'' of this Poisson manifold, that is
a submanifold on which the Poisson structure induces  the symplectic structure. It is the symplectic
structure that we mean in the title.

The symplectic manifold in question is closely related with the well known, ``standard''
symplectic manifold, the orbit of coadjoint representation of the Lie group.
It is well investigated class of manifolds, and the problem of their canonical parametrization dates
back to Archimedes who found a symplectomorphism between sphere, that
can be considered as an orbit of $\operatorname{SU}(2)$,
and a circumscribed cylinder.

In the present article  we construct an explicit 
 birational symplectic isomorphism between
the symplectic leaf of the space of Fuchs  equations and a Cartesian product
of the coadjoint orbits (for the version of the method
 for $2\times 2$ matrices see \cite{Babich}). It solves the problem of the canonical parametrization
 in the sense that any set of canonical coordinates on a coadjoint orbit
gives us a set of desired canonical coordinates of the
leaf.

The suggested method does not depend on the type of normal Jordan form
of matrices $A^{(n)}$. The only difference between not-diagonalizable and
diagonalizable cases is following.

In the diagonalizable case the method gives the
{\em isomorphism} between some Zariski open domain of the
leaf and the Cartesian product of the orbits, but in the
non-diagonalizable case it gives {\em the birational isomorphism} only.
The same effect we have in a
 Painlev\'e non-diagonalizable case  $2\times 2$ matrices too.
 When we parameterize
 the equations by a degenerate matrix
 $$\left(
     \begin{array}{cc}
     pq & q \\
       -p^2q &-pq  \\
     \end{array}
   \right),
 $$
 we add a divisor $q=0, p\in \CC$ that does NOT belong to the orbit of
 $\left(\begin{array}{cc}
  0 & 1 \\
  0 & 0 \\
  \end{array}
  \right)$,
so new divisor
that does not belong to the orbit arise.

\section{Spaces $\tilde M$, $\tilde M|_{\scriptscriptstyle \Sigma=0}$ and $M$.}
Let $A^{(1)}, \dots , A^{(N)}\in \operatorname{gl}^N(m)$
be a set of $N\ge 3$ matrices $m\times m$.\\
Let $\chi_{
 n}=\det (\lambda \operatorname{I}-A^{(n)})$ be a characteristic
polynomial of  $A^{(n)}$.

\begin{Restriction}\label{restriction}
We assume that the matrices $A^{(n)}$ have the following property:
 all eigen subspaces of every $A^{(n)}$ are one-dimensional.
 \end{Restriction}
The Restriction implies that all invariant factors (see \cite{Gelfand}, 	
another name -- {\em invariant
polynomials}, see \cite{Gantmacher}) except a determinant are equal to unit.

\begin{note}
A conjugacy  class of matrices under the Restriction~\ref{restriction}  is well defined
by its characteristic polynomial.
\end{note}

Denote the conjugacy  class  of matrices with one-dimensional eigenspaces
and characteristic polynomial
$\chi$  by $\mathcal{O}(\chi)$:
$$\mathcal{O}(\chi):=\bigcup_{g\in \GLm}g^{-1}Ag, \ \text{ where }
\chi:=\det (\lambda \operatorname{I}-A)
$$

\begin{note}
The restriction implies that $A^{(n)}$ commutes with  polynomials on $A^{(n)}$ only, and
 the orbit $\mathcal{O}(\chi_n)$ is $m(m-1)$-dimensional.
\end{note}

It is well known that $\mathcal{O}(\chi_n)$ is $m(m-1)$-dimensional
symplectic space, let us denote it symplectic form by $\omega_n$.

Let us consider $N$ polynomials $\chi_n, \ n=1, \dots , N$ of the $m$'s order with unit
leading coefficients. Let us denote by
$(\tilde M, \tilde \omega)=(\tilde M, \tilde \omega)_{\chi_1, \dots, \chi_N}$ a symplectic space
$$\mathcal{O}(\chi_1)\times \dots \times \mathcal{O}(\chi_N)=:\tilde M, \ \ \
\tilde \omega :=\omega_1+\dots +\omega_N.
$$

Such a space is parameterized by $N\times m$ numbers, these are coefficients of $\chi_n, \ n=1, \dots, N$.
Elements of $\tilde M$ are the sets
$A^{(1)}, \dots , A^{(N)}$. We denote such a set by $A^{(\bar n)}: \
A^{(\bar n)}\in \tilde M$.

Let us consider a diagonal action of $\GLm$ on $\tilde M$:
$$g^{-1}A^{(\bar n)}g:=g^{-1}A^{(1)}g, \dots ,g^{-1}A^{(N)}g\in \tilde M, \ \ \ g\in\GLm.
$$
It is well known
that it is a Poisson action (see \cite{Arnold}, Appendix 5), and a momentum
map is
$$A^{(\bar n)}\longrightarrow \sum_{n=1}^NA^{(n)}.
$$
Let us denote a zero level of the momentum by $\tilde M|_{\scriptscriptstyle \Sigma=0}$:
$$A^{(\bar n)}\in \tilde M|_{\scriptscriptstyle \Sigma=0}=
\mathcal{O}(\chi_1)\times \dots \times \mathcal{O}(\chi_N)|_{\scriptscriptstyle \Sigma=0}
\ \Longleftrightarrow \
\sum_{n=1}^NA^{(n)}=0.
$$
By a classical Marsden-Weinstein theorem (see \cite{MW,Arnold,RS}) the factor
with respect to the
action
of corresponding group
of a level of momentum
is a symplectic space, let us denote it by $M$:
$$M:=\tilde M|_{\scriptscriptstyle \Sigma=0}/\GLm =
\mathcal{O}(\chi_1)\times \dots \times \mathcal{O}(\chi_N)|_{\scriptscriptstyle \Sigma=0}/
\GLm .
$$

A point of $M$ we denote by $(\!(A^{(\bar n)})\!)$, it is an equivalence class
of sets $A^{(\bar n)}$:
$$A^{(\bar n)}\sim {A'}^{(\bar n)} \ \ \ \Longleftrightarrow \ \ \ \exists g\in \GLm : \
A^{(\bar n)}=g^{-1}{A'}^{(\bar n)}g$$

\bigskip

The goal of this paper is to construct explicitly a symplectic birational isomorphism
between  $M$ and {\em a standard} symplectic space, that is a Cartesian product of the orbits
$\mathcal{O}(\chi)$.

\section{Subset $\tilde D\stackrel{\pi}{\to}D$ of $\tilde M|_{\scriptscriptstyle \Sigma=0} \stackrel{\pi}{\to} M$}

Let us consider a fiber bundle $\tilde M|_{\scriptscriptstyle \Sigma=0} \stackrel{\pi}{\to} M$.
The Hamiltonian reduction theory states that a form $\tilde \omega|_{\scriptscriptstyle \Sigma=0}$,
that is a restriction of $\tilde \omega$ on the submanifold $\sum A^{(n)}=0$, takes
the same values on all vectors $\tilde \xi \in \mathrm{T}\tilde M$
with the same projection on $\mathrm{T}M$:
$$
\pi \tilde\xi_1=\pi \tilde\xi_2:=\xi, \ \
\pi \tilde\eta_1=\pi \tilde\eta_2:=\eta
\ \ \Longrightarrow \ \
\tilde \omega (\tilde\xi_1,\tilde\eta_1)=
\tilde \omega (\tilde\xi_2,\tilde\eta_2)=: \omega (\xi , \eta ).
$$
It means that $\tilde \omega$ is
in the image of $\pi^*$, of the pullback of $\pi$. It implies
that on the image of $\pi$ the form $\omega$ is
well defined by an equality $\tilde\omega =\pi^*\omega$.
The Hamiltonian reduction theory guarantees  that $\omega$ is symplectic, i.e. non-degenerate
and closed.

A regular way to construct (local coordinate) functions on the factor-manifold
 $M$ is to specify
a (local) section $\sigma_{\scriptscriptstyle D}$ of the bundle
$$
\sigma_{\scriptscriptstyle D}: D 
\to
\tilde M|_{\scriptscriptstyle \Sigma=0}, \ \
\pi\sigma_{\scriptscriptstyle D}=id, \ \text{ where } D\subset M,
$$
and restrict on the $\sigma_{\scriptscriptstyle D}(D)\subset \tilde M$
functions defined on $\tilde M$, in particular -- matrix elements $A^{(n)}_{ij}$, they form
a standard set of coordinate functions on $\tilde M$.

Practically we specify a domain $\tilde D\subset \tilde M$, and restrict
the fibre bundle $\tilde M|_{\scriptscriptstyle \Sigma=0} \stackrel{\pi}{\to} M$ on it:
${\tilde D} \stackrel{\pi}{\to} D:=\pi (\tilde D)$.
There are rational sections of this bundle. We define such a section as a graph,
a subset $\sigma_{\scriptscriptstyle D}(D)\subset \tilde D\subset \tilde M|_{\scriptscriptstyle \Sigma=0}$
that intersects all fibres of $\tilde D\stackrel{\pi}{\to}D$ exactly once.
 The Definition-Notation\ref{domain_tilde_D} below is a definition-description of these submanifolds.
Let us realize this programm.
\bigskip

\begin{Def-Not}\label{domain_tilde_D}
By $\tilde D:=\tilde D (N,N-1,N-2)$ we denote such a subset of
$\tilde M|_{\scriptscriptstyle \Sigma=0}\supset \tilde D$ that
for sets $A^{(\bar n)}$ from it there exists such a $g\in\GLm $ that
\begin{enumerate}
  \item $g^{-1}A^{(N-1)}g$ is upper-triangular,
  \item $g^{-1}A^{(N-2)}g$ is lower-triangular
  \item  $\sum_{k=1}^m(g^{-1}A^{(N)}g)_{ik}=\sum_{k=1}^m(g^{-1}A^{(N)}g)_{jk}$ $\forall i,j$.
\end{enumerate}
\end{Def-Not}

To discuss these three conditions, consider an action of $\GLm$ on the set of frames
of $\CC^m$. There is one-to-one correspondence (faithful representation)
between changes of a fixed basis
$\mathfrak{(E)}\stackrel{def}{=}(\vec e_1, \dots , \vec e_m)$ of $\CC^m$,
and elements $g$ of $\GLm$: $\mathfrak{(E)}\to \mathfrak{(E')}=\mathfrak{(E)}g$.
Elements  of algebra $\glm$ can be considered as linear transformations
of this $\CC^m$; changes of the basis of $\CC^m$
induce the adjoint representation of the group on its algebra by similarity transformations.

In this interpretation Definition-Notation \ref{domain_tilde_D} means that there is such a basis $\mathfrak{(E)}$
of $\CC^m$ that three specified matrices $A^{(N-1)},A^{(N-2)}$ and $A^{(N)}$
have special forms. Consider $A^{(N-1)}, A^{(N-2)}, A^{(N)}$ in the basis $\mathfrak{(E)}$.

\bigskip

First of all, let us notice that $A^{(N-1)}$ and $A^{(N-2)}$  are triangular, consequently their diagonal
elements are eigenvalues. The specification of the basis with necessary properties (1),(2) gives
some ordering of their eigenvalues $\lambda_1^{(N-1)}, \dots , \lambda_m^{(N-1)}=:\lambda_{\vec k}^{(N-1)}$,
$\lambda_1^{(N-2)}, \dots , \lambda_m^{(N-2)}=:\lambda_{\vec k}^{(N-2)}$.

Let us consider the last property (3). It means that one of eigenvectors of $A^{(N)}$ has
all the components equal to each other, let us add its eigenvalue (denote it by $\lambda^{(N)}$)
that is a common value of the sums in (3),
to the set of discreet parameters -- the orders of eigenvalues.

Let us denote by $\mathfrak{(E)}( \lambda^{(N)}; \lambda_{\vec k}^{(N-1)},\lambda_{\vec k}^{(N-2)})$ the basis
in which diagonal elements of $A^{(N-1)},A^{(N-2)}$ have preassigned orders of eigenvalues,
and the sum of every raw of $A^{(N)}$ is equal to $\lambda^{(N)}$ that is the assigned eigenvalue of $A^{(N)}$.

These orderings are the discreet parameters of our construction. In a generic case we can choose
$\lambda^{(N)}$ and orderings of eigenvalues of $A^{(N-1)}, A^{(N-2)}$ in an arbitrary way, in special points
of $\tilde D$ some combinations of $\lambda$'s can not be realized, but the defining property
of $\tilde D$ is the following: {\em there is at least one such a basis $\mathfrak{(E)}$}.

\section{Upper- and lower- normal Jordan forms related to the
assigned ordering of the diagonal elements}
Let us fix parameters $ \lambda^{(N)}, \lambda_{\vec k}^{(N-1)},\lambda_{\vec k}^{(N-2)}$ in some a way
and consider a sub-domain $\tilde D(N,N-1,N-2;
 \lambda^{(N)}; \lambda_{\vec k}^{(N-1)},\lambda_{\vec k}^{(N-2)})\subset \tilde D (N,N-1,N-2)$, where
the basis $\mathfrak{(E)}( \lambda^{(N)}; \lambda_{\vec k}^{(N-1)},\lambda_{\vec k}^{(N-2)})$ exists.

We say that matrix $A^{(N-1)}_{J+}$
has upper-Jordan form {\em related to the ordering $\lambda_{\vec k}^{(N-1)}$
of its eigenvalues}, if it is upper-triangular with assigned order of the diagonal entries;
all non-diagonal entries are zero except units that correspond to the Jordan chains of generalized eigenvectors.

The only difference from the standard Jordan normal form  is that the eigenvectors and the
generalized eigenvectors that form the basis of $\CC^m$ in which the matrix has the standard
Jordan form are rearranged in the special order in accordance with the given order of the diagonal elements.

In the same way we define a lower-Jordan form $A^{(N-2)}_{J-}$
{\em related to the ordering $\lambda_{\vec k}^{(N-2)}$
of the eigenvalues} of $A^{(N-2)}$.

The difference between upper- and lower- forms in the directions of Jordan chains.

The following simple fact from the matrix theory is very important for us.

\begin{prop}
Any upper-(lower-)triangular matrix can be transformed
into the upper-(lower-)Jordan normal form with the same diagonal via similarity transformation
by some upper-(lower-)triangular matrix.

If all the eigenspaces of the matrix are one-dimensional (Restriction \ref{restriction}), all such similarity
transformations differ from each other by upper-(lower-)triangular factor,
that is an arbitrary diagonal for the diagonalizable matrix.

\end{prop}

\proof

The first statement is evident. The second one is true  because if we have only one
eigenvector corresponding to the eigenvalue, the only freedom in
the procedure of constructing of Jordan basis is a choice of the lead vector of the
Jordan chain. It gives the triangular transformation of the set of
base vectors in the invariant subspace
corresponding to this eigenvalue.
\qed

\section{Gauss decomposition and function $\mathfrak{(E)}(\cdot)$}
\bigskip
Let $A$ and $B$ be any matrices.
Let us denote a basis in which $A$ has
upper-Jordan form  related to a fixed ordering of eigenvalues by $\mathfrak{(E_+)}$,
and let us denote a basis in which $B$ has lower-Jordan form
related to a fixed ordering of eigenvalues by $\mathfrak{(E_-)}$.
Let us denote an element of $\GLm$ that transforms $\mathfrak{(E_+)}$ to $\mathfrak{(E_-)}$ by $\Phi_{+-}$:

$$\mathfrak{(E_+)}\Phi_{+-}= \mathfrak{(E_-)}, \ \ \ \Phi_{+-}\in \GLm .
$$

\begin{thm}\label{Gauss_decomposition}
The following two statements are equivalent:
\begin{itemize}
  \item[I.]\label{item_a} Matrix $\Phi_{+-}$ admits the Gauss decomposition
  $$\Phi_{+-}=\Phi_+\Phi_-
  $$
  on the upper- and lower- triangular $\Phi_{+}$ and $\Phi_{-}$.
  \item[II.]\label{item_b} There is such a basis $\mathfrak{(E)}$, in which $A$ is upper-triangular and $B$ is lower-triangular.
  \end{itemize}
\end{thm}

\proof
If $\Phi_+\Phi_-$ is the Gauss decomposition of $\Phi_{+-}$, the desired basis $\mathfrak{(E)}$ is:
$$\mathfrak{(E)}=\mathfrak{(E_+)}\Phi_+=\mathfrak{(E_-)}\Phi_-^{-1},
$$
so 
``I'' implies 
``II''.

 Otherwise, let $\mathfrak{(E)}$ be a basis in which $A$ is upper-triangular, $B$ is lower-triangular.
 A transformation of an upper-(lower-)triangular matrix to the corresponding
 upper-(lower-)Jordan form can be made by a triangular matrix.

 Let us denote the triangular matrices
 that transform $A$ and $B$ to the corresponding (upper- and lower-) normal Jordan form by
 $\Phi_+^{-1}$ and $\Phi_-$.
 They are the matrices
 that give us the Gauss decomposition of
 the transformation connecting $\mathfrak{(E_+)}$ and $\mathfrak{(E_-)}$.
 \qed

\begin{thm}\label{Gauss_uniqueness}
Let $A$ and $B$ be any matrices without two-dimensional eigenspaces,
and let condition ``II'' of the Theorem~\ref{Gauss_decomposition} fulfil. Then

$\mathfrak{(E)}$ is uniquely defined up to
 an arbitrary diagonal transformation $\delta$: $\mathfrak{(E)}\sim \mathfrak{(E)}\delta $.
\end{thm}
\noindent In other words the directions of the vectors of basis  in question are uniquely defined.

\proof

Upper-(lower-) Jordan normal form commutes with matrices of a special type
only, that is triangular in the  case when only one
Jordan block corresponds to any eigenvalue (see \cite{Gantmacher}).
That is our case, consequently
bases  $\mathfrak{(E_\pm)}$ are defined up to the corresponding
(upper- and lower-) triangular transformations
$\delta^{\scriptscriptstyle \Delta}_{\pm}$ (of special kind):
$\mathfrak{(E'_+)}=\mathfrak{(E_+)}\delta^{\scriptscriptstyle \Delta}_+$,
$\mathfrak{(E'_-)}=\mathfrak{(E_-)}\delta^{\scriptscriptstyle \Delta}_-$.

Gauss decomposition of a given $\Phi_{+-}$ is unique up to a diagonal factor. The ambiguity
of the choices of $\mathfrak{(E_\pm)}=\mathfrak{(E'_\pm)}\delta^{\scriptscriptstyle \Delta}_\pm$
gives the following ambiguity of $\Phi_{+-}$:
  $$\Phi'_{+-}=\delta^{\scriptscriptstyle \Delta}_+\Phi_{+-}(\delta^{\scriptscriptstyle \Delta}_-)^{-1},
  $$
  it has evident decomposition on the product of triangular factors:
  $$\Phi'_{+-}=\delta^{\scriptscriptstyle \Delta}_+\Phi_{+-}
  (\delta^{\scriptscriptstyle \Delta}_-)^{-1}=(\delta^{\scriptscriptstyle \Delta}_+
  \Phi_{+})(\Phi_{-}(\delta^{\scriptscriptstyle \Delta})_-^{-1})
  =:\Phi'_{+}\Phi'_{-}.$$
  By the uniqueness of the Gauss decomposition  $\Phi'_{+}$ and $\Phi'_{-}$ are defined up to diagonal factors --
  right for $\Phi'_{+}$ and left for $\Phi'_{-}$, that is the source of diagonal ambiguity of $\mathfrak{(E)}$.

  The ambiguity of $\mathfrak{(E_\pm)} : \ \mathfrak{(E'_\pm)}=
  \mathfrak{(E_\pm)}(\delta^{\scriptscriptstyle \Delta}_+)^{\pm 1}$ does not
  effect on the basis $\mathfrak{(E)}$:
  $$\mathfrak{(E)}=\mathfrak{(E_+)}\Phi_+=\mathfrak{(E'_+)}\delta^{\scriptscriptstyle \Delta}_+
  \Phi_+=\mathfrak{(E'_+)}\Phi'_+=\mathfrak{(E)},
  $$
  it proves the present Theorem.
  \qed

 \bigskip
Consider the Definition-Notation~\ref{domain_tilde_D}. It states that for the sets $A^{(\bar n)}$ from the
domain $\tilde D$ it is possible to choose such a basis $\mathfrak{(E)}$ that $A^{(N-1)}$ will be
upper-triangular, and $A^{(N-2)}$ will be lower-triangular. By the Theorem~\ref{Gauss_uniqueness}
basis $\mathfrak{(E)}$ is determined the up to the right diagonal factor.

Let us consider the third
condition  of Definition-Notation~\ref{domain_tilde_D}. It is equivalent  to the
equality $A^{(N)}\vec 1=\lambda^{(N)}\vec 1$, where $\vec 1$ is a column of units,
and $\lambda^{(N)}$ is a common value of the sums. Changes of basis change coordinates of eigenvectors:
$\mathfrak{(E)}=\mathfrak{(E')}\delta \ \ \to \ \ \vec f=\delta \vec {f'}$. Consequently the conditions of
the Definition-Notation~\ref{domain_tilde_D} determine the basis $\mathfrak{(E)}$
up to a common scalar factor. It is the base of our construction in a direct and a
figurative sense. Let us reformulate it.

Let us denote by $\mathbf{P}(\text{`` bases of $\CC^m$ ''})$ the projectivization of the set of frames.
Different representatives of one class are bases connected by a scalar factor. In any basis
from one equivalence class all matrices $A$ have the same matrix elements.
We have constructed the sections of a projectivised frame-bundle over  $\tilde D$, that are the
single-valued functions $\mathfrak{(E)}$ with values in  $\mathbf{P}(\text{`` bases of $\CC^m$ ''})$.
Let us denote the set of data $(N,N-1,N-2;\lambda^{(N)};\lambda_{\vec k}^{(N-1)},\lambda_{\vec k}^{(N-2)})$
by $(\cdot)$ for short.

\begin{thm}\label{rationality}
Function $\mathfrak{(E)}(\cdot)$: $\tilde D (\cdot)  \ \to \ \mathbf{P}(\text{`` bases of $\CC^m$ ''})$
is well defined, and all the components $\mathfrak{(E)}_i^j$ of the vectors of the basis $\mathfrak{(E)}$
are rational functions of matrix elements
$A^{(n)}_{ij}, n=N,N-1,N-2; i,j,k =1, \dots ,m$.
\end{thm}

\proof
Determing eigenvectors and generalized eigenvectors by a given Jordan
normal form is a problem of solving the systems of linear equations, the same for
Gauss decomposing of matrices collected from these eigenvectors and generalized eigenvectors.
These are rational operations.
\qed

\begin{thm}
$\tilde D(\cdot)$ and $D(\cdot)$ are open sets in a Zariski topology.
\end{thm}
\proof
A complement to $\tilde D$ is defined by a system
of algebraic equations -- vanishing of upper-left minors of $\Phi_{+-}$,
that is the condition of the impossibility of Gauss decomposing (see \cite{Gantmacher}),
and vanishing of any component of the eigenvector $\vec f$ of $A^{(N)}$.

Complement to $D$
is defined by the same equations, because the conditions of the Definition-Notation~\ref{domain_tilde_D}
are invariant with respect to the action of $\GLm$.
\qed

\bigskip
Let us restrict the fibre bundle $\tilde M|_{\scriptscriptstyle \Sigma=0} \stackrel{\pi}{\to} M$ on
$\tilde D(\cdot )\subset \tilde M|_{\scriptscriptstyle \Sigma=0}$: $\tilde D (\cdot) \stackrel{\pi}{\to} D(\cdot)$.
From now on we somehow  fix the parameters $(N,N-1,N-2;\lambda^{(N)};\lambda_{\vec k}^{(N-1)},\lambda_{\vec k}^{(N-2)})$
and will not write $(\cdot)$ any more.

\begin{prop}
$\tilde \omega|_{\scriptscriptstyle \tilde D}=
\pi^*\omega|_{\scriptscriptstyle D}$,
and 2-form $\omega|_{\scriptscriptstyle D}$ is a symplectic form on $D$
\end{prop}
\proof
Making the restriction we exclude from the consideration only several submanifolds  of smaller
dimensions.
\qed

\bigskip
We have  constructed the function $\mathfrak{(E)}:D\to \mathbf{P}(\text{`` bases of $\CC^m$ ''})$, that
induces a rational section $\sigma$ of fibre-bundle $\tilde D \stackrel{\pi}{\to} D$:
$$\sigma(
\ (\!(
A^{(\bar n)}
)\!)
 \ )=
 A^{(1)}_\sigma, \dots , A^{(N)}_\sigma=A^{(\bar n)}_\sigma,
$$
where $A^{(\bar n)}_\sigma$ is a representative of $(\!( A^{(\bar n)})\!)$
calculated in the basis $\mathfrak{(E)}$.
In other words, it is such a representative of the conjugacy class that
has the triangular forms  of $A^{(N-1)}, A^{(N-2)}$ with assigned orders of the diagonal elements
and the assigned value of sums of all rows of $A^{N}$.

Section $\sigma$ depends on discreet parameters:
$N,N-1,N-2;\lambda^{(N)};\lambda_{\vec k}^{(N-1)},\lambda_{\vec k}^{(N-2)}$, but we will not write them  again.

 \section{Projection ${\hat \pi}^{(N)}_{\lambda^{(N)}}$ on the orbit $\mathcal{O}$ of the smaller dimension}

Let us denote a natural projections  on the Cartesian factors $\mathcal{O}(\chi_n)$ by $\pi^{(n)}$.
These projections  are defined for all points of $
 \glm \times \dots \times \glm$:
$$\pi^{(n)}: \ \ A^{(1)}, \dots , A^{(N)} 
\longrightarrow
A^{(n)}\in \mathcal{O}(\chi_n).
$$
Let us consider a submanifold $\sigma (D)\subset \tilde D$.  For its points we
will define a projection
${\hat \pi}^{(N)}_{\lambda^{(N)}}: \ \mathcal{O}(\chi_N)\to \mathcal{O} (\chi_N/\lambda-\lambda^{N})$,
where the orbit $\mathcal{O} (\chi_N/\lambda-\lambda^{N})$ corresponds to the
polynomial $\chi_N/\lambda-\lambda^{N}$. Its degree is less than the degree of $\chi_N$ by a unit:
$\mathcal{O} (\chi_N/\lambda-\lambda^{N})\subset \operatorname{gl}(m-1)$.

\bigskip
Let us do it. All matrices from $\mathcal{O} (\chi_N)\cap \sigma (D)$ have a fixed eigenvector $\vec 1$,
we take it as an $m$'th  basis vectors of $\CC^m$: $\mathfrak{(\widehat{E})}:=\mathfrak{(E)}\Xi$, where
$\Xi$
is the following constant matrix
$$
\Xi:=\left(
  \begin{array}{cccccc}
    1      & 0            & \cdots & 0 &0       & 1      \\
    0      & 1             & \cdots & 0 &0       & 1      \\
    \vdots & \vdots & \ddots  & \vdots & \vdots & \vdots \\
    0      & 0&\cdots       &      1 & 0      &1       \\
    0      & 0&\cdots   &      0 & 1      &1       \\
    0      & 0&\cdots   &      0 & 0      &1       \\
  \end{array}
\right), \ \Xi^{-1}=
\left(
  \begin{array}{cccccr}
    1      & 0            & \cdots & 0 &0       & -1      \\
    0      & 1             & \cdots & 0 &0       & -1      \\
    \vdots & \vdots & \ddots  & \vdots & \vdots & \vdots \\
    0      & 0&\cdots       &      1 & 0      &-1       \\
    0      & 0&\cdots   &      0 & 1      &-1       \\
    0      & 0&\cdots   &      0 & 0      &1       \\
  \end{array}
\right)
$$
In the basis $\mathfrak{(\widehat{E})}$ matrix  $A^{(N)}$ takes a block-triangular form:
\begin{equation}\label{block-triangle}
\Xi^{-1}A^{(N)}\Xi =:
\left(
                         \begin{array}{cc}
                           \hat A & \vec 0 \\
                            a^T & \lambda^{(N)} \\
                         \end{array}
                       \right),
\end{equation}
where by $\hat A$ we denote a matrix whose  characteristic polynomial is $\chi_N/\lambda-\lambda^{(N)}$;
$a^T$ is an $m-1$-dimensional vector-row, it is the last row of $A^{(N)}$ without the $m$'s
element: $a^T_k=A^{(N)}_{mk}$, $k\ne m$.

\begin{Def-Not}\label{projection}
Let us denote by ${\hat \pi}^{(N)}_{\lambda^N}$ a map
$${\hat \pi}^{(N)}_{\lambda^N}: \ \
A^{(1)}, \dots , A^{(N)} \longrightarrow \hat A,
$$
where $\hat A$ is $m-1\times m-1$ upper-left-corner block of $\Xi^{-1}A^{(N)}\Xi$.
\end{Def-Not}

\begin{prop}
${\hat \pi}^{(N)}_{\lambda^N}(A^{(\bar n)})$ is
a projection on the orbit $\mathcal{O}(\chi_N/\lambda-\lambda^{N})$.
\end{prop}
\proof
By a construction the characteristic polynomial of $\hat A$
is $\chi_N/\lambda-\lambda^{N}$, so it is sufficient to prove that
all eigenspaces of $\hat A$ are one-dimensional.

\bigskip
Let $\lambda^{(N)}=0$. It is not a restriction because we can subtract $\lambda^{(N)}\mathbf{I}$ from $A^{(N)}$,
it  will not change the dimensions of the eigenspaces of the block. Assume that there are
two different (not proportional) eigenvectors $f_1,f_2$ of $\hat A$ that correspond to one $\lambda$.

If $\lambda\ne 0$,  vectors
$(f_i, <a^Tf_i>/\lambda)^T, \ \ i=1,2$ are two different eigenvectors of $\Xi^{-1}A^{(N)}\Xi$
corresponding to one $\lambda$ that contradicts Restriction~\ref{restriction}.
Let $\lambda=0$. If, say, $<a^Tf_1>=0$ then $(f_1, 0)^T$ is the eigenvector
of $\Xi^{-1}A^{(N)}\Xi$ in addition to $(\vec 0,1)^T$. If $<a^Tf_1>\ne 0\ne <a^Tf_1>$,
the second eigenvector corresponding to $\lambda=0$ is $(<a^Tf_2>f_1-<a^Tf_1>f_2,0)$.
It is not a zero because of the linear independence of $f_i$.
\qed

\begin{Def-Not}
Let us denote by $\widehat \pi$ a following projection of $\sigma (D)$ on the Cartesian product of the orbits:
$
\sigma (D)\longrightarrow
\mathcal{O}(\chi_N/\lambda-\lambda^{(N)})
\times
\mathcal{O}(\chi_1)\times \dots \times \mathcal{O}(\chi_{N-3})$,
$$
\widehat \pi (A^{(\bar n)}):= {\hat \pi}^{(N)}_{\lambda^N} (A^{(\bar n)}), \pi^{(1)}(A^{(\bar n)}),
\dots, \pi^{(N-3)}(A^{(\bar n)}).
$$
\end{Def-Not}
The product of the orbits is a symplectic space, consequently we construct
the rational map between two symplectic spaces. The first space is $D$, that is an open dense domain of
$M:=
\mathcal{O}(\chi_1)\times \dots
\times \mathcal{O}(\chi_N)|_{\scriptscriptstyle \Sigma=0}/\GLm \supset D$,
and the second space is the
 Cartesian product of the orbits $\mathcal{O}$.
 The main theorem of the present article is

\begin{thm}[]\label{main}
The map $ \widehat{ \pi}\sigma$ is a symplectic birational isomorphism;
in a case of  diagonalizable $A^{(n)}$'s, that is a case of a general position,
$\widehat \pi\sigma$ is a symplectic isomorphism that is birational
between $D$ and the product of the orbits.
\end{thm}

Before starting the proof let me remind  that by the Hamiltonian reduction
theory the $\tilde \omega|_{\scriptscriptstyle \Sigma=0}$ on
$\tilde M|_{\scriptscriptstyle \Sigma=0}$ is degenerate and its
 value is the same on all vectors with the same projection on $M$.
We can {\em reformulate} this statement in the following way:\\

Let us consider {\em any} local section that is a (local) isomorphism on its image:
$\sigma': M\to\tilde M|_{\scriptscriptstyle \Sigma=0}, \ \pi\sigma'=id$.
Let us restrict $\tilde \omega|_{\scriptscriptstyle \Sigma=0}$ on the image.
We get
$(\tilde \omega|_{\scriptscriptstyle \Sigma=0})|_{\sigma'(M)}:=\tilde{\tilde{ \omega}}$;
$\sigma'$ is the (local) {\em isomorphism}, consequently its inverse ${\sigma'}^{-1}$ exists
and define a desired 2-form $({\sigma'}^{-1})^*\tilde{\tilde{ \omega}}$
on (a domain of) $M$, that is the pullback of $\tilde{\tilde{ \omega}}$ by the ${\sigma'}^{-1}$.

Generically this form (of cause!) depends on the choice of the section $\sigma'$.
It is the main result  of the theory of the Hamiltonian reduction ---
{\em to specify the conditions}  (a Poisson action of a group,
level of a momentum map etc.) {\em in such a way} that 2-form
$({\sigma'}^{-1})^*\tilde{\tilde{ \omega}}$
{\em would not} depend on the choice of the section $\sigma'$.
It is our freedom that we can choose $\sigma'$ as we like.
It follows from the presented speculations that the value of the sum
\begin{equation}\label{omega_sum}
\omega=\omega_1+\dots +\omega_N
\end{equation}
 does {\em not} depend on the choice
of the submanifold $\sigma (D)$
on which we restrict the sum.
We choose it {\em in  the most comfortable} way --
actually it is the Definition-Notation~\ref{domain_tilde_D}.
Let us prove the Theorem now.

\proof Consider the projection $\hat \pi^{(N)}_{\lambda^N}$ of the $\mathcal{O}(\chi_N)$ on
$\mathcal{O}(\chi_N/\lambda-\lambda_N)$.
Let us denote by
${{{\hat{ \pi}}^{(N)*}_{\lambda^N}}}\hat \omega$
the pullback of the standard symplectic form
$\hat{ \omega}$ on the orbit $\mathcal{O}(\chi_N/\lambda-\lambda_N)$ by this projection.

We have two forms defined on $\mathcal{O}(\chi_N)$, the standard
symplectic form $\omega_{N}$ and the pullback in question.
They can not coincide because $\omega_{N}$ is symplectic but
 the pullback is  degenerate
-- the dimension of $\mathcal{O}(\chi_N/\lambda-\lambda_N)$
is smaller than the dimension of $\mathcal{O}(\chi_N)$.

Let us consider the projection $\hat \pi^{(N)}_{\lambda^N}$ of $D$ on
$\mathcal{O}(\chi_N)$. It is not surjective.
Its image is a submanifold $\hat \pi^{(N)}_{\lambda^N}(D)$ of a smaller dimension.
Consider the two mentioned forms {\em restricted on the submanifold}
$\hat \pi^{(N)}_{\lambda^N}(D)\subset \mathcal{O}(\chi_N)$.


\begin{lem}
$$({\hat \pi}^{(N)*}_{\lambda^N}\hat \omega)|_{\hat \pi_{\lambda^N}^{(N)}(D)} =
(\omega_N)|_{\hat \pi^{(N)}_{\lambda^N}(D)}
$$
\end{lem}
\proof
Let us print the expression~\cite{Hitchin} for the standard Lie-Poisson form on an orbit:
$$\omega_{LP}(\xi,\eta)=- \operatorname{tr} U_\xi \dot A_\eta=\operatorname{tr} U_\eta \dot A_\xi,
$$
where $\xi,\eta\in T_A\mathcal{O}$ are two tangent vectors to an orbit $\mathcal{O}$ in a point
$A\in \mathcal{O}$:
$$\xi=[A,U_\xi]=\left.\frac{d}{dt}\right|_{t=0}A_{\xi}(t)=:\dot A_\xi, \
\eta=[A,U_\eta]=\left.\frac{d}{dt}\right|_{t=0}A_{\eta}(t)=:\dot A_\eta,
\ \ A_{\xi}(0)=A_{\eta}(0)=A.
$$
All matrices $A^{(N)}$ in the special basis $\mathfrak{(E)}$
(actually they form  the $\hat \pi^{(N)}_{\lambda^N}(D)$)
have the constant eigenvector $\vec 1$, and after the similarity transformation
by the constant matrix $\Xi$ take the block-triangle form (\ref{block-triangle}).
The printed expression for the Lie-Poisson form is invariant with respect to
the constant similarity transformations and having
been calculated for the block-diagonal matrix with the fixed
diagonal element $\lambda^{(N)}$ gives the desired --- it is just the same expression,
where we should write the
first $m-1\times m-1$ upper-left-diagonal blocks of  all matrices
$\dot A_\xi, \dot A_\eta, U_\xi, U_\eta$ transformed by $\Xi$,
that is the projection $\hat \pi^{(N)}_{\lambda^N}$.
\begin{flushright} 
Lemma has been proved $\Box$
\end{flushright}

We have proved that the term $\omega_{N}$, in the sum (\ref{omega_sum}),  can be replaced by
${\hat \pi}^{(N)*}_{\lambda^N}\hat \omega$, i.e.
we can calculate the contribution of $A^{(N)}$ after its projecting
on $\mathcal{O}(\chi_N/\lambda-\lambda^{(N)})$.

Let us consider the summands $\omega_{N-1}$ and $\omega_{N-2}$.
By construction the
images $\pi^{(N-1)}\sigma(A^{(\bar n)})$ and $\pi^{(N-2)}\sigma(A^{\bar n})$
belong to Lagrangian
submanifolds of triangular matrices, consequently corresponding summands vanish
and we get the announced
{\em rational symplectic projection of $D\subset M$ on the product of the $N-2$ orbits}.

To finish the proof we should construct a rational inverse transformation.

\bigskip
We have $N-3$ matrices $A^{(n)}\in \mathcal{O}(\chi_n)$, the
diagonal entries
$\lambda_{\vec n}^{(N-1)}$,  $\lambda_{\vec n}^{(N-2)}$
 of the triangular $A^{(N-1)}$, $A^{(N-1)}$, eigenvalue $\lambda^{(N)}$ of $A^{(N)}$ and
  matrix $\hat A\in \mathcal{O}(\chi_N/\lambda-\lambda_N)$ as the initial data.
  We have to reconstruct $A^{(N)}$, $A^{(N-1)}$, $A^{(N-2)}$ using this data
 and the equation $\sum_{n=1}^NA^{(n)}=0$.

 To do this explicitly we introduce some notations

 \bigskip
For any matrix $A\in \mathrm{gl}(m)$ let us denote its projections on
the diagonal, under-diagonal and over-diagonal
subalgebras by $A_=$, $A_>$ and $A_<$:
$$(A_=)_{ij}=A_{ii}, \ i=j, \ \
(A_=)_{ij}=0, \ i\ne j$$
$$(A_>)_{ij}=A_{ij}, \ i>j, \ \
(A_>)_{ij}=0, \ i\le j, $$
$$
(A_<)_{ij}=A_{ij}, \ i<j, \ \
(A_<)_{ij}=0, \ i\ge j$$

First of all let us construct the diagonal part of $A^{(N)}$.
All the diagonal entries of $A^{(n)}, \ n<N$ are given
 --- matrices $A^{(N-1)}$ and $A^{(N-2)}$ are diagonal with eigenvalues on their diagonals,
consequently we have to put
$$A_{=}^{(N)}:=-\sum_{n=1}^{N-1}A_{=}^{(n)}.
$$
Now let us construct $A^{(N)}$. It is a matrix on the orbit $\mathcal{O}(\chi_N)$
with given projection $\hat A$ and given diagonal.
Matrix $\Xi$ that defines the projection $A^{(N)}\to \hat A$ has such simple
structure that we can easily write the answer.
By the given diagonal we can specify
the last row of $A$ in such a way that $\hat A$ will be the assigned $m-1\times m-1$ matrix.

To present an explicit formula for $A^{(N)}$ let us introduce
the operation $\hat A \to\stackrel{\circ}{\hat A} $,
that makes $m\times m$ matrix $\stackrel{\circ}{\hat A}$
from any $m-1\times m-1$ matrix $\hat A$ by adding the last zero row and the last zero column.
Let us denote this embedding of $\operatorname{gl} (m-1) \to \glm $ by a small circle
over the name of a matrix: $\hat A\to \stackrel{\circ}{\hat A}$.

Some more notations.
Let us denote
$\vec{ \mathbf 1}\in \mathbb{C}^m$, $\vec{ \mathbf 1} \otimes \vec{ \mathbf 1}\in \mathrm{gl}(m)$,
 $(0\dots 01) \otimes \vec{ g}\in \mathrm{gl}(m)$, where $\vec g$ is any vector:

$$(\vec{ \mathbf 1})_i=1; \
 (\vec{ \mathbf 1} \otimes \vec{ \mathbf 1})_{ij}=1 \forall i,j; \
 ((0\dots 01) \otimes \vec{ g})_{ij}=0, \  j\ne m, \ ((0\dots 01) \otimes \vec{ g})_{im}=g_i.$$
It can be easily  verified that
$$
A^{(N)}=\stackrel{\circ}{\hat{A}}+(\vec{ \mathbf 1} \otimes \vec{ \mathbf 1})
({A_{=}^{(N)}-\stackrel{\circ}{\hat{A}}}_{=})-
(0\dots 01) \otimes (\stackrel{\circ}{\hat{A}}\vec{ \mathbf 1} ).$$

Here $(0\dots 01) \otimes
(\stackrel{\circ}{\hat{A}}\vec{ \mathbf 1} )$ is the product of the row $(0\dots 01)$
on the column (it is $\vec g$) that is the product of $\stackrel{\circ}{\hat{A}}$ and the
column $\vec{ \mathbf 1}$;
$({A_{=}^{(N)}-\stackrel{\circ}{\hat{A}}}_{=})$ is a diagonal matrix $m\times m$.

Matrix $A^{(N)}$ and diagonal entries of all other matrices are determined now. Finally
$$
A_{<}^{(N-1)}:=-\sum_{\substack{n=1\\ n\ne N-1}}^NA_{<}^{(n)},\ \
A_{>}^{(N-2)}:=-\sum_{\substack{n=1\\  n\ne N-2}}^NA_{>}^{(n)}, \ \
A_{>}^{(N-1)}=A_{<}^{(N-2)}=0\in \glm .$$

\qed

\end{document}